\documentclass[draft]{aipproc}

\layoutstyle{6x9}

\SetInternalRegister\hbadness{8000} % pseudo latin isn't breaking very well :-)

\newcommand\doingARLO[2][]{%
  \ifx\mmref\undefined #1\else #2\fi
}

\begin{document}

\title 
{Shell-Model Calculations with Realistic Effective Interactions}


\author{L. Coraggio}{
  address={Dipartimento di Scienze Fisiche, Universit\`a di Napoli Federico II
and Istituto Nazionale di Fisica Nucleare, Complesso Universitario di Monte 
S. Angelo, Via Cintia, 80126 Napoli, Italy},
}

\iftrue
\author{A. Covello}{
  address={Dipartimento di Scienze Fisiche, Universit\`a di Napoli Federico II
and Istituto Nazionale di Fisica Nucleare, Complesso Universitario di Monte 
S. Angelo, Via Cintia, 80126 Napoli, Italy},
}

\author{A. Gargano}{
  address={Dipartimento di Scienze Fisiche, Universit\`a di Napoli Federico II
and Istituto Nazionale di Fisica Nucleare, Complesso Universitario di Monte 
S. Angelo, Via Cintia, 80126 Napoli, Italy},
}

\author{N. Itaco}{
  address={Dipartimento di Scienze Fisiche, Universit\`a di Napoli Federico II
and Istituto Nazionale di Fisica Nucleare, Complesso Universitario di Monte 
S. Angelo, Via Cintia, 80126 Napoli, Italy},
}

\author{T. T. S. Kuo}{
  address={Department of Physics, State University of New York at Stony Brook, 
Stony Brook, New York 11794},
}
\fi

% \copyrightholder{Acoustical Scociety of America}
\copyrightyear  {2001}

\begin{abstract}
In this paper, we present some results of shell-model calculations
employing effective interactions derived from the CD-Bonn free
nucleon-nucleon potential.
These concern  $^{18}$O, $^{134}$Te, and $^{210}$Po, and are part of a
comprehensive study of nuclei around doubly closed shells.
Comparison of the calculated results with experimental data
illustrates the degree of accuracy of modern realistic shell-model
calculations.
\end{abstract}

%\date{\today}

\maketitle

\section{Introduction}

A basic ingredient in nuclear shell-model calculations is the model-space 
effective interaction.
The most popular approach to the determination of $V_{\rm eff}$ relies on 
the use of empirical effective interactions, which leads in general to a 
very successful description of nuclear structure properties. 
While this approach is certainly of great practical value for the 
interpretation of experimental data, it is not satisfactory from a 
first-principle point of view. 
In fact, a fundamental goal of nuclear structure theory is to understand the 
properties of complex nuclei in terms of the nucleon-nucleon ($NN$) 
interaction. 
This implies the derivation of $V_{\rm eff}$ from the free $NN$ potential. 

Over the last decade there has been substantial progress towards a 
microscopic approach to nuclear structure calculations starting from a free 
$NN$ potential. 
This has concerned both the two basic ingredients which come into play in this 
approach, namely the $NN$ potential and the many-body methods for deriving the 
model-space effective interaction.

As regards the first point, several $NN$ potentials have been constructed, 
which give a very accurate description of the $NN$ scattering 
data and are suitable for application in nuclear structure.
These are the Nijmegen potentials \cite{stoks94}, the Argonne ${\rm
V_{18}}$ potential \cite{wir95}, and the very recent high-precision CD-Bonn
potential \cite{mach01} which, according to the analysis made in
\cite{mach01}, gives the most accurate reproduction of the presently 
available $NN$ data.

As for the second point, an accurate calculation of the Brueckner $G$-matrix 
has become feasible while the so-called $\hat Q$-box folded-diagram series for 
the effective interaction $V_{\rm eff}$ can be summed up to all orders using 
iterative methods. 
An outline of the derivation of $V_{\rm eff}$ will be given in section 2. 

Based on these improvements, in recent years there has been a renewed
interest in shell-model calculations employing effective interactions derived 
from the free $NN$ potential. 
In this context, a key issue is how accurate a description of nuclear
structure properties can be provided by these ``realistic'' effective
interactions.

On this problem, we have concentrated our efforts during the last few years 
[4-9]. 
We have focused attention on nuclei with few valence particles 
or holes, since they provide the best testing ground for the basic 
ingredients of shell-model calculations, especially as regards the matrix 
elements of the effective $NN$ interaction.

It is our aim here to present some selected results of our
calculations, which have been obtained by making use of the CD-Bonn 
$NN$ potential, and the main conclusions of our study. 

\section{The shell-model effective interaction}

Let us now outline our derivation of $V_{\rm eff}$.
Because of the strong repulsive core, which is a feature common to all modern 
$NN$ potentials, the model-space $G$ matrix corresponding to the chosen 
$V_{NN}$ must be calculated first.
The $G$ matrix is defined \cite{krenc76}  by the integral equation:

\begin{equation}
G(\omega)=V+VQ_2 \frac{1}{\omega-Q_2TQ_2}Q_2G(\omega),
\end{equation}

\noindent
where $V$ represents the $NN$ potential, $T$ denotes the two-nucleon kinetic
energy, and $\omega$ is an energy variable (the so-called starting energy).
The two-body Pauli exclusion operator $Q_2$ prevents double counting, namely
the intermediate states allowed for $G$ must be outside the chosen model
space.
Thus the Pauli operator $Q_2$ is dependent on the model space, and so
is the $G$ matrix.
The operator $Q_2$ is specified, as discussed in Ref. \cite{krenc76}, by a set
of three numbers $(n_1,n_2,n_3)$ each representing a single-particle orbital.
Note that in Eq.(1) the Pauli exclusion operator $Q_2$ is defined in
terms of harmonic oscillator wave functions while plane-wave functions are 
employed for the intermediate states of the $G$ matrix.  

Our procedure for calculating the $G$ matrix goes as follows. 
We first calculate the free $G$ matrix $G_F$ in a proton-neutron
representation, $G_F$ being defined by
\begin{equation}
G_F=V+V\frac{1}{e}G_F,
\end{equation}
with $e\equiv (\omega - T)$. Note that $G_F$ does not contain the 
Pauli exclusion operator and hence its calculation is relatively convenient. 
Then we calculate the Pauli correction term \cite{krenc76,tsai72},
\begin{equation}
\Delta G=-G_F\frac{1}{e}P_2\frac{1}{P_2(\frac{1}{e}+\frac{1}{e}G_F\frac{1}{e}
)P_2}P_2
\frac{1}{e}G_F,
\end{equation}
where $P_2 = 1 - Q_2$, separately for protons and for neutrons. 
Finally, the full $G$ matrix as defined
by Eq.(1) is obtained as
\begin{equation}
G=G_F +\Delta G.
\end{equation}

For the harmonic oscillator parameter $~\hbar \omega$ we use the value 
obtained from the expression $~\hbar \omega=(45A^{-1/3}-25A^{-2/3})$ MeV.

Using the above $G$ matrix we then calculate the irreducible vertex
function $\hat{Q}$-box, which is composed of irreducible valence-linked
$G$-matrix diagrams through second order.
These are precisely the seven first- and second-order diagrams considered by
Shurpin {\em et al.} \cite{shurp83}.
The effective interaction can be written in operator form as

\begin{equation}
V_{eff} = \hat{Q} - \hat{Q'} \int \hat{Q} + \hat{Q'} \int \hat{Q} \int
\hat{Q} - \hat{Q'} \int \hat{Q} \int \hat{Q} \int \hat{Q} + ~...~~,
\end{equation}

\noindent
where $\hat{Q}$ is the $\hat{Q}$-box, and the integral sign represents a
generalized folding operation \cite{kuo80}.
$\hat{Q'}$ is obtained from $\hat{Q}$ by removing terms of first order in the
reaction matrix $G$.
After the $\hat{Q}$-box is calculated, the energy-independent $V_{\rm eff}$ is
obtained by summing up the folded-diagram series of Eq.(5) to all orders
using the Lee-Suzuki iteration method \cite{suzuki80}.
This last step can be performed in an essentially exact way for a given
$\hat{Q}$-box.

\section{Results}

To illustrate the degree of accuracy to which realistic shell-model 
calculations can describe the spectroscopic properties of nuclei near
closed shells, we report here some results we have obtained for 
light-, medium- and heavy-mass nuclei having two valence particles.
They have been obtained by using the OXBASH shell-model code \cite{brown}.

\begin{table}[h]
\begin{tabular}{rrr}
\hline
 & \tablehead{2}{c}{} {{\em E}(MeV)} \\
\tablehead{1}{r}{}{$J^{\pi}$}
  & \tablehead{1}{r}{}{Exp.}
  & \tablehead{1}{r}{}{Calc.} \\
\hline
$0^+$ & 0.000 &  0.000 \\
$2^+$ & 1.982 &  1.824 \\
$4^+$ & 3.555 &  3.269 \\
$0^+$ & 3.634 &  3.720 \\
$2^+$ & 3.920 &  3.889 \\
\hline
\end{tabular}
\caption{Experimental and calculated low-energy levels in $^{18}$O.}
\label{tab:a}
\end{table}

As already mentioned in the Introduction, we have made use of effective 
interactions derived from the CD-Bonn $NN$ potential. 
Regarding the single-particle energies, we have taken them from the 
experimental spectra of the corresponding single-particle valence nuclei. 

\begin{table}[ht]
\begin{tabular}{rrr}
\hline
 & \tablehead{2}{c}{} {{\em E}(MeV)} \\
\tablehead{1}{r}{b}{$J^{\pi}$}
  & \tablehead{1}{r}{b}{Exp.}
  & \tablehead{1}{r}{b}{Calc.} \\
\hline
$0^+$ & 0.000 &  0.000 \\
$2^+$ & 1.279 &  1.188 \\
$4^+$ & 1.576 &  1.472 \\
$6^+$ & 1.691 &  1.605 \\
$6^+$ & 2.398 &  2.249 \\
$2^+$ & 2.462 &  2.438 \\
\hline
\end{tabular}
\caption{Experimental and calculated low-energy levels in $^{134}$Te.}
\label{tab:b}
\end{table}

\begin{table}[ht]
\begin{tabular}{rrr}
\hline
 & \tablehead{2}{c}{} {{\em E}(MeV)} \\
 \tablehead{1}{r}{b}{$J^{\pi}$}
  & \tablehead{1}{r}{b}{Exp.}
  & \tablehead{1}{r}{b}{Calc.} \\
\hline
$0^+$ & 0.000 &  0.000 \\
$2^+$ & 1.181 &  1.076 \\
$4^+$ & 1.427 &  1.342 \\
$6^+$ & 1.473 &  1.435 \\
$8^+$ & 1.557 &  1.498 \\
$8^+$ & 2.188 &  2.123 \\
$2^+$ & 2.290 &  2.250 \\
\hline
\end{tabular}
\caption{Experimental and calculated low-energy levels in $^{210}$Po.}
\label{tab:c}
\end{table}

In Tables 1-3 we compare the experimental \cite{nndc} and calculated 
low-energy levels of $^{18}$O, $^{134}$Te, and $^{210}$Po. 
We see that the observed excitation energies in all the three nuclei are 
very well reproduced by our calculations, the discrepancy being less than 
150 keV for most of the states. 

We refer the reader to papers \cite{andr97,cor99}, where we have made 
use of the Bonn-A $NN$ potential \cite{mach87}, for a more complete 
analysis of the spectroscopic properties of $^{134}$Te and $^{210}$Po. 

\section{Summary}

In this paper, we have presented some recent results of realistic 
shell-model calculations, which are part of a comprehensive study
aimed to assess the role of realistic effective interactions in
nuclear structure theory.

The main conclusion of our study is that this kind of calculations is 
able to provide a quantitative description of nuclear structure
properties. 
It should be noted, however, that most of our calculations 
have until now concerned nuclei with identical valence nucleons.
A careful test of the $T=0$ matrix elements is of course equally important. 
In this regard, we may mention that in the study \cite{andr99} of the 
doubly odd nucleus $^{132}$Sb we have obtained results which are as good as 
those regarding like nucleon systems. 
Along the same lines we are currently studying other nuclei with both 
neutrons and protons outside closed shells.

\begin{theacknowledgments}
This work was supported in part by the Italian Ministero dell'Universit\`a 
e della Ricerca Scientifica e Tecnologica (MURST) and by th U.S. DOE Grant No. 
DEFG02-88ER40388. 
N.I. would like to thank the European Social Fund for financial support.
\end{theacknowledgments}

% choose bibtex style depending on layout style and options used in
% sample:

\doingARLO[\bibliographystyle{aipproc}]
          {\ifthenelse{\equal{\AIPcitestyleselect}{num}}
             {\bibliographystyle{arlonum}}
             {\bibliographystyle{arlobib}}
          }
%\bibliography{sample}

\end{document}